# Pressure sensing with Zero Group Velocity Lamb modes in self-supported a-SiC/c-ZnO membranes

C. Caliendo[*,1], M. Hamidullah[1]

[1]Institute for Photonics and Nanotechnologies, IFN-CNR, Via Cineto Romano 42, 00156 Rome, Italy

*Corresponding author; email: cinzia.caliendo@cnr.it

**Abstract.** The propagation of the Lamb modes along a-SiC/ZnO thin supported composite structures was simulated for different ZnO and a-SiC layer thicknesses and electrical boundary conditions. The phase and group velocity, the field profile and the electroacoustic coupling coefficient dispersion curves of the Lamb modes travelling along the composite plate were calculated for different layers thicknesses. Zero group velocity (ZGV) points were identified which group velocity vanishes while the phase velocity remains finite, at specific layers thickness values. ZGV resonators (ZGVRs) were designed that consist in only one interdigital transducer (IDT) and no grating reflectors at its sides. The finite element method analysis was performed to investigate the strain, stress and internal pressure the a-SiC/ZnO plate experiences when subjected to an external uniform differential pressure in the 1 to 10 kPa range. The ZGVR pressure sensitivity, i.e. the relative frequency shift per unit pressure change, was found to be mostly affected by the change in the membrane thickness induced by the pressure. A pressure sensitivity of 9 ppm/kPa, in the 4 to 10 kPa range, was predicted for the a-SiC(1μm)/ZnO(1μm) ZGV-based pressure sensor. The feasibility of high-frequency micro-pressure sensors based on a-SiC and ZnO thin film technology was demonstrated by the present simulation study.

## 1. Introduction

The acoustic waves that propagate in finite-thickness *isotropic* homogeneous plates are divided into two classes: Lamb waves and shear horizontal (SH) plate waves. These waves are decoupled and can be distinguished into symmetric and anti-symmetric modes, $S_n$ and $A_n$, where n is the mode order; an infinite number of $S_n$ and $A_n$ modes exist in a plate. The Lamb modes are elliptically polarized in that they show two particle displacement components, $U_1$ and $U_3$, parallel and orthogonal to the wave propagation direction *x*. $U_1$ lies in the plane that contains the wave vector *k*: its depth profile is symmetric with respect to the mid-plane of the plate for symmetric modes while it is antisymmetric for the antisymmetric modes. $U_3$, the shear vertical component, is perpendicular to the plate surface: its depth profile is antisymmetric to the mid-plane of the plate for the symmetric modes, while it is symmetric for the antisymmetric ones [1, 2]. The SH plate modes are horizontally polarized and have only one particle displacement component, $U_2$, that lies in the plane that contains the wave vector *k* and is orthogonal to *k*. Guided waves travelling in a homogeneous *anisotropic* plate have three particle displacement components ($U_1$, $U_2$ and $U_3$.) since the Lamb waves and the SH waves are generally coupled [3]. If the plane of propagation coincides with a plane of symmetry of the material, then Lamb modes are decoupled from the SH modes, and symmetric and anti-symmetric modes can be still distinguished. The characteristics of Lamb waves propagating in non-homogeneous *composite* plates (i.e., bi-layered composite plates) are more complex than in homogeneous isotropic plates as the waveguide symmetry with respect to the mid plane of the plate is lost and the modes can no more be identified as symmetric or antisymmetric: the modes are generically distinguished by a number in the order in which they appear along the frequency axis. All the modes are dispersive as their velocity depends on the plate thickness-to-wavelength ratio H/λ, and some modes have group and phase velocities with opposite signs. For some branches of the angular frequency dispersion curves, ω vs $k$, a strong resonance occurs at the frequency minimum corresponding to a zero-group velocity (ZGV) Lamb mode: this stationary non-propagating mode is characterized by a vanishing group velocity $v_{gr} = \partial\omega/\partial k$ combined with a non-zero wave number $k$ [4–8]. The ZGV points appear in the frequency spectrum of both monolayer (isotropic and anisotropic) and multilayer plates. Depending on the plate material type and crystallographic orientation, in addition to modes with a *single* ZGV point, some modes exhibit *double* and even *multiple* such points [9] as their dispersion curve undergoes multiple changes in the sign of its slope. If the composite waveguide consists in a piezoelectric layer (such as ZnO or AlN) and a non-piezoelectric layer, the Lamb wave propagation can be excited and detected by use of interdigitated transducers (IDTs), as for the surface acoustic waves (SAWs). The IDTs can be either positioned onto the free surface of the piezoelectric layer or buried under the piezoelectric layer, thus allowing the exploitation of different electroacoustic coupling configurations.

ZGV resonators (ZGVRs) are associated with an intrinsic energy concentration beneath the IDT. Due to the vanishing group velocity, the acoustic energy cannot be carried away from the IDT, leading to a stationary non-propagating mode. They can be fabricated with a technology simpler than that required by the surface acoustic wave resonators (SAWRs) and Lamb wave resonators (LWRs). In the SAWRs the acoustic energy is entirely confined to the substrate surface: the IDT and the two grating reflectors positioned at its sides define the boundary of the planar resonant cavity. The most severe limitation of the SAWRs are the low velocity and Q factor. LWRs are based on two well-known configurations. The *grating-type* [10] configuration employs one or two IDTs and two or three reflective gratings, similar to the one- or two-port SAWRs [11]: the waves launched by the IDTs are reflected back by the gratings to reduce the energy loss [12]. The *free edges* [13] configuration includes only the IDT and no reflectors: the waves propagate in the piezoelectric thin plate until they reflect from the suspended free edges of the thin plate. Another example of LWR is the ZGVR: it requires only one IDT and a properly selected plate thickness-to-wavelength value [14, 15]. At the ZGV point, the energy cannot be carried away from the IDT, leading to a strong resonance of the plate. The ZGVRs have dimensions smaller than those of the LWRs or SAWRs as they consist in only one interdigital transducer and no grating reflectors at the IDT's sides. The ZGVRs are capable of solving the low frequency and Q limitation faced by the SAWRs as the phase velocity of the ZGV modes is much larger than the SAW velocity.

Non-piezoelectric amorphous SiC (a-SiC) shows excellent properties such as a high acoustic wave velocity, resistance to chemicals, high hardness and compatibility with the integrated circuit technology as it can be deposited by a r.f. magnetron sputtering system [16-18] at 200°C onto Si(100) substrates from a sintered SiC target. Piezoelectric wurtzite ZnO thin film technology has been widely used for many years for the fabrication of SAW devices onto non-piezoelectric substrates, such as silicon, glass, and sapphire, to name just a few. When the piezoelectric ZnO film is grown onto high-velocity materials, such as diamond or SiC, it is a promising candidate for high frequency, high sensitivity micro sensors [19]. A bi-layered a-SiC/ZnO composite thin plate, few micrometers thick, can be obtained by standard technological processes, such as the magnetron sputtering growth of the a-SiC and ZnO layers onto a Si(100) wafer, and the backside Si/a-SiC/ZnO micro-machining process for the fabrication of suspended membranes. In this case the a-SiC film plays the role of a back-etching stop layer, allowing the release of the a-SiC/ZnO suspended membrane.

This paper provides a simulation study of the ZGV modes in a-SiC/ZnO waveguides for different layers thicknesses. The ZGVR pressure sensitivity, i.e. the relative resonant frequency shift per unit pressure change, has been calculated under external uniform differential pressure, in the 1 to 10 kPa range. The finite element method (FEM) analysis has been performed to investigate the ZGVR strain, stress and internal pressure, thus allowing to get a further insight into the ZGVR-based pressure sensitivity. The feasibility of high-frequency micro-pressure sensor based on a-SiC and ZnO thin film technology was demonstrated by the present simulation study. A pressure sensitivity (9 ppm/kPa in the 4 to 10 kPa range) higher or at least comparable to that of SAWRs-based sensors [20-22] was predicted for the 1-1 ZGV2-based pressure sensor.

## 2. LAMB MODES IN a-SiC PLATES

The phase and group velocity dispersion curves of the Lamb modes propagating along the bare amorphous SiC (a-SiC) thin plate were calculated using the Disperse code [12] (matrix method software DISPERSE) and are shown in Figures 1a-b for an a-SiC plate with thickness H = 5 μm; the a-SiC elastic constants ($c_{11}$ = 375.3 and $c_{12}$ = 112.5 GPa ) and the mass density ($\rho$ = 3079 kg/m$^3$) were extracted from reference 23.

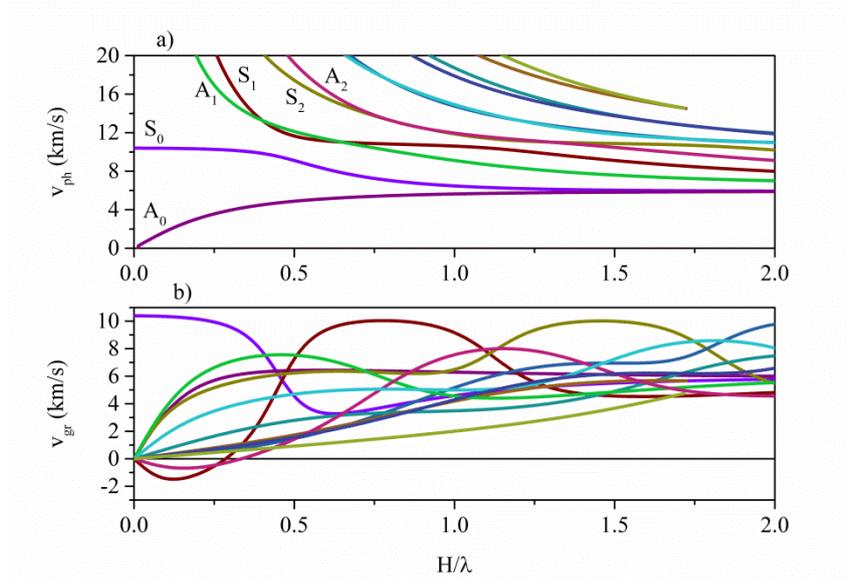

**Figure 1.** The phase and group velocity, $v_{ph}$ and $v_{gr}$, vs thickness-to-wavelength ratio H/λ curves of the Lamb modes travelling in a bare a-SiC plate, 5 μm thick.

As it can be seen in Figure 1, the first symmetric and second anti-symmetric modes, $S_1$ and $A_2$, are characterized by a ZGV point to which a non-zero wave number (non-zero phase velocity) corresponds. At this specific plate thickness-to-wavelength ratio H/λ the acoustic energy does not propagate in the waveguide (as the $v_{gr}$ is null), resulting in sharp resonance effects, as opposed to the plate thickness resonances which associated with $k = 0$. The $v_{gr}$ of the $S_1$ mode vanishes at frequency $f_0$ = 1033 MHz (phase velocity $v_{ph}$ =1.8075 km/s); the $v_{gr}$ of the $A_2$ mode vanishes at frequency $f_0$= 1902.8 MHz ($v_{ph}$ =2.824 km/s). Figures 2a and b show the mode shape and the power flow density of the $S_1$ and $A_2$ ZGV points; the three quantities have been reduced to dimensionless quantities. The ZGV power flow, the integral of the power flow density in the wave propagation direction (positive *x* direction), over the plate thickness, $\int_{-H/2}^{H/2} P_x$, is zero, as opposed to the cases of the backward-wave motion ($v_{gr}$ and $v_{ph}$ have opposite sign) and forward-wave motion ($v_{gr}$ and $v_{ph}$ have equal sign) which correspond a negative and positive power flow, respectively. The middle of the membrane was chosen to be the zero depth (rather than the surface) in order to emphasize the symmetry with respect to the plate midplane.

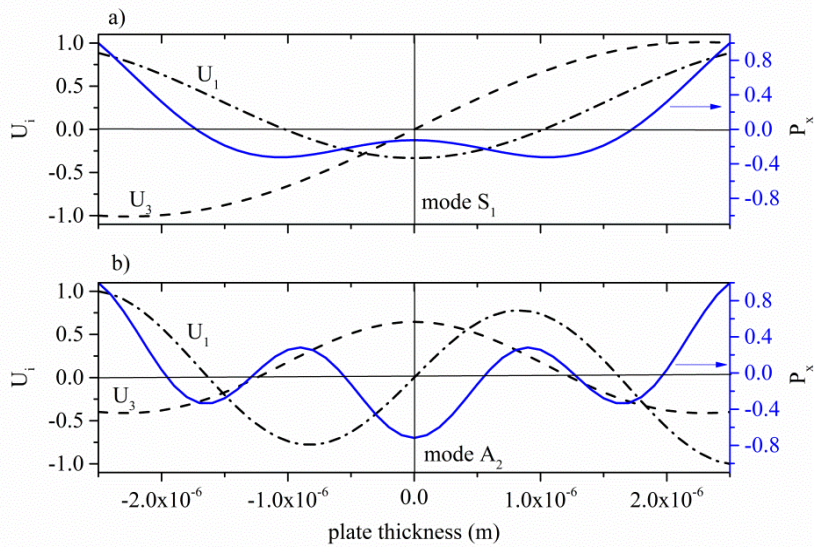

**Figure 2.** The power flow density $P_x$ and the mode shape of the ZGV points corresponding to a) the $S_1$ mode and b) the $A_2$ mode of an a-SiC plate, 5 μm thick. The middle of the membrane was chosen to be the zero depth.

For the $S_1$ mode, $U_1$ is symmetric around the mid-plane of the plate while $U_3$ is antisymmetric; for the $A_2$ mode, $U_3$ is symmetric while $U_1$ is antisymmetric with respect to the mid-plane of the plate. $U_1$ and $U_3$ are normalized to the $U_1$ value at the plate surface [24, 25].
Since a-SiC is not piezoelectric, a thin piezoelectric film is required to cover the a-SiC surface in order to excite the Lamb modes propagation by means of IDTs.

### 3. LAMB MODES IN ZnO/a-SiC PLATES
The phase and group velocity dispersion curves of the ZnO/a-SiC composite plates were calculated by the software DISPERSE [26] for different layers thicknesses. The w-ZnO material constants were extracted from reference 27; as the Disperse software doesn't account for the ZnO piezoelectric constants, then its database was provided with the ZnO stiffened elastic constants calculated with a Matlab routine. The ZnO/a-SiC composite structure brakes the mid-plane symmetry, thus the symmetric and antisymmetric nature of the modes is hardly distinguished, except for the two fundamental modes, named quasi-$S_0$ (q$S_0$) and quasi-$A_0$ (q$A_0$); the higher order modes are labeled with a progressive number as well as the corresponding ZGV points. Different ZnO/a-SiC composite plates were modelled with variable a-SiC and ZnO layers thickness in the 1 to 5 μm range. These plates will be labelled hereafter with two numbers: the first corresponds to the ZnO layer thickness in μm, and the second corresponds to the thickness of the a-SiC layer. As an example, figures 3a-b show the phase velocity dispersion curves of the Lamb modes travelling in the ZnO/a-SiC 2.5-5 and 5-2.5 composite plates: it can be noticed that the number of the propagating modes and their velocity are highly affected by the two layers thickness.

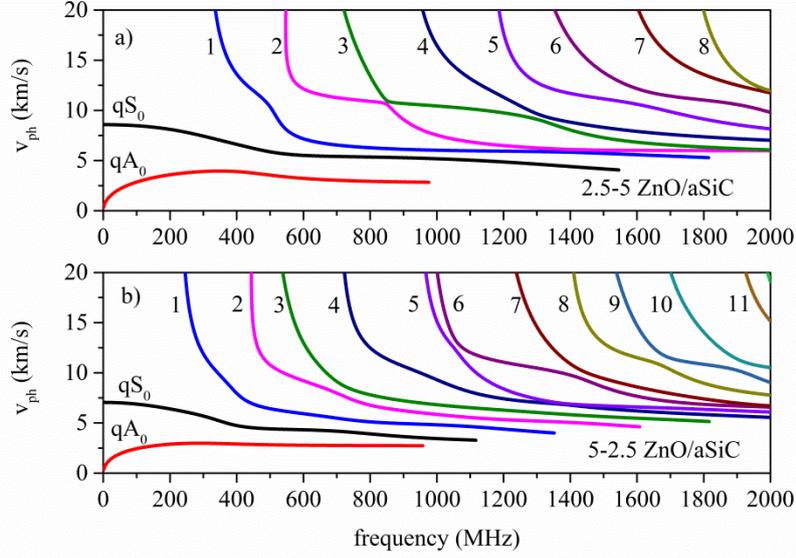

**Figure 3.** The phase velocity dispersion curves of the Lamb modes traveling along the a) 2.5-5 and b) 5-2.5 ZnO/a-SiC composite plates. The four coupling configurations of the ZGV resonator.

In the studied frequency range (from few MHz up to 2 GHz), it was found that the 1-5 and 2-5 composite plates exhibit only one ZGV point corresponding to the mode 2 (namely the quasi-$S_1$ mode, $qS_1$) at frequencies $f_0$ = 763.633 and 604.668 MHz, hereafter named ZGV2. The 3-5 and 4-5 composite plates exhibit three ZGV points corresponding to mode 2 (at frequencies $f_0$ = 499.535 and 424.938 MHz), mode 5 ($f_0$ = 1059.72 and 900.896 MHz), and mode 8 ($f_0$ = 1590.61 and 1334.08 MHz), named hereafter ZGV2, ZGV5 and ZGV8. The 5-5 composite plate exhibits up to five ZGV points corresponding to mode 2, mode 5, mode 8, mode 11 and mode 14, named ZGV2, ZGV5, ZGV8, ZGV11 and ZGV14, at $f_0$ = 369.378, 781.106, 1157.21, 1570.64 and 1954.31 MHz, respectively. As an example, figures 4a-b show the group velocity $v_{gr}$ dispersion curves of the modes 2 and 5, for different ZnO layer thicknesses and fixed a-SiC layer thickness (5 μm). The abscissa is the total plate thickness-to-wavelength ratio $H_{tot}/\lambda = (h_{a-SiC} + h_{ZnO})/\lambda$.

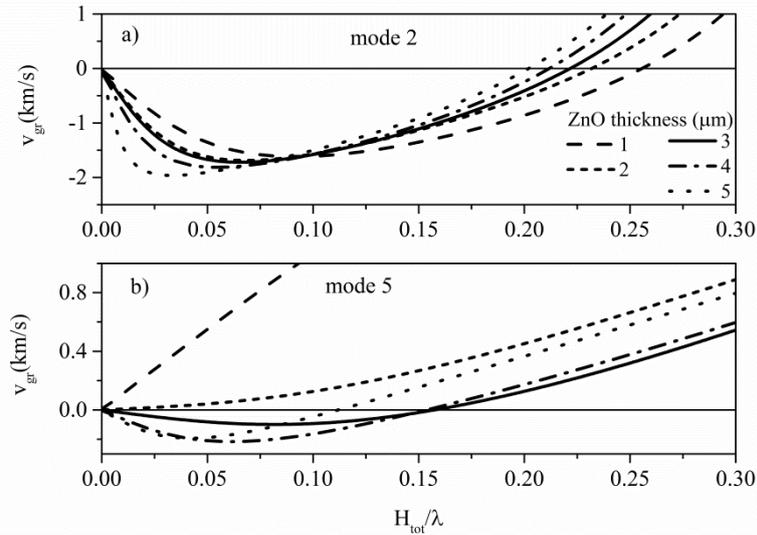

**Figure 4.** The group velocity vs $H_{tot}/\lambda$ of the a) mode 2 and b) mode 5 of the ZnO/a-SiC plate; the a-SiC thickness is fixed (5 μm) while the ZnO thickness is the running parameter.

With increasing the piezoelectric layer thickness from 1 to 5μm, the abscissa of the ZGV2 decreases as the corresponding wavelength increases and thus the ZGV2 resonant frequency $f_0 = v_{ph}/\lambda$ moves toward lower values. The dispersion curves of the mode 5 do not cross the zero group velocity axis for the 1-5 and 2-5 plates, as opposed to the thicker plates.

The mode shape of the ZGV2, ZGV5 and ZGV8 for the 3-5 composite plate is shown in figures 5a-c as an example. The middle of the composite membrane, 8 μm thick, was chosen to be the zero depth in order to emphasize the plate asymmetry with respect to its midplane. The acoustic field profile of the ZGV2 still shows some peculiarities typical of the first anti-symmetric mode $S_1$. The field profile of the ZGV5 and of the ZGV8, have the characteristic shape of a higher order mode but it is artificial to identify a similarity with a symmetric or anti-symmetric mode.

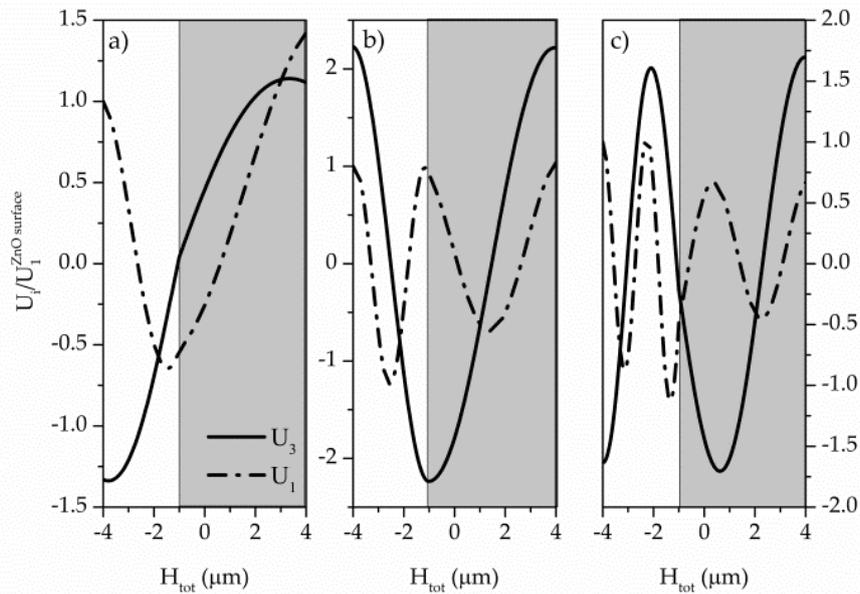

**Figure 5.** The mode shape of the a) ZGV2, b) ZGV5 and c) ZGV8 of the 3-5 structure; the grey area defines the a-SiC thickness. The middle of the composite membrane, 8 μm thick, represents the zero depth.

Finite element method (FEM) simulations have been carried out by using COMSOL Multiphysics 5.2 to explore the field shape of the ZGVRs in the composite waveguide. The simulations accounted for five Al IDT finger pairs, 0.1 μm thick and $\lambda/4$ wide, located onto the free surface of the ZnO layer: the terminal (1V) and ground electrical boundary conditions were applied at the top surface of the interdigitated electrodes alternately. Two perfectly matched layers (PML), each one wavelength wide, were applied on the left and right side of the a-SiC/ZnO plate, in order to model a domain with open boundaries through which the wave pass without undergoing any reflection; the traction free boundary conditions were applied to the top and bottom sides of the composite plate. The total length of the studied cell is $20\cdot\lambda$, including the two PMLs. The maximum and minimum mesh size were $\lambda/10$ and $\lambda/100$. The ZnO was assumed to have an elastic loss $\tan\delta = 0.002$. As an example, figures 6a-c show the absolute total displacement of three zero group velocity points (ZGV2, ZGV5 and ZGV8) belonging to the 3-5 plate, and corresponding to $\lambda$ = 36, 49 and 35 μm respectively. As it can be seen, the displacement is confined only in the region underneath the IDT.

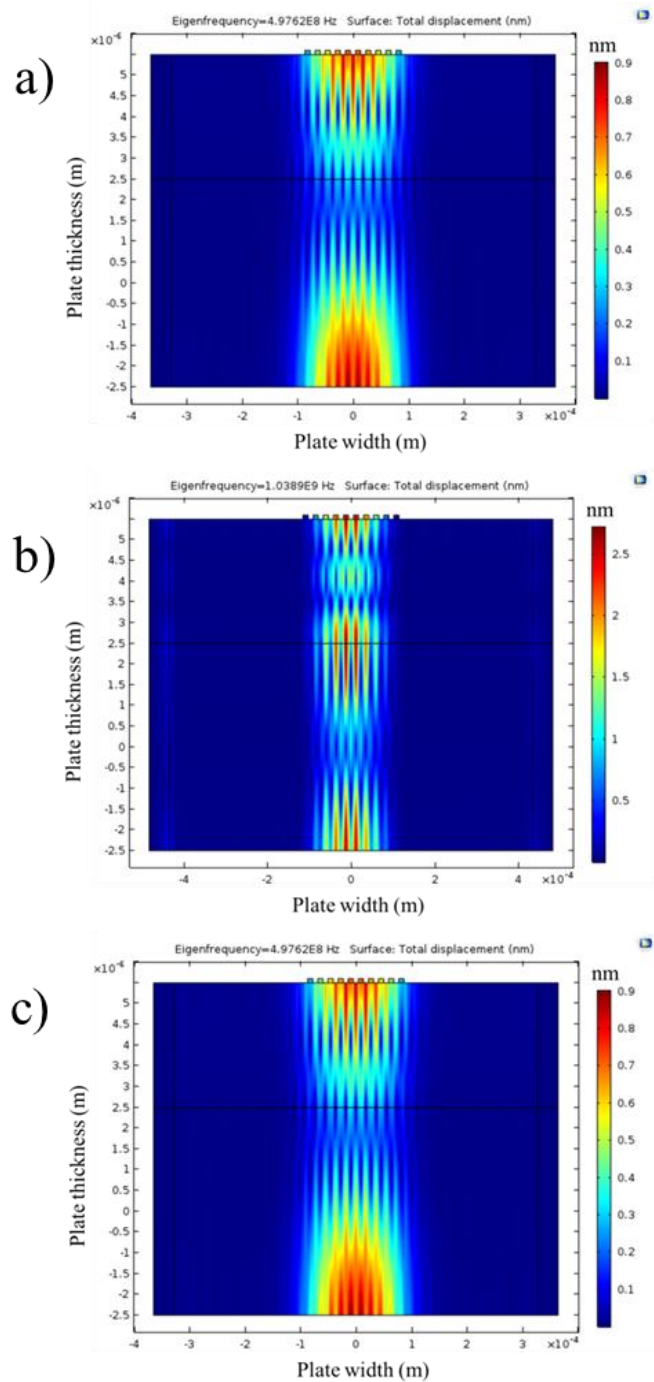

**Figure 6.** The absolute surface total displacement of the (a) ZGV2, (b) ZGV5, and (c) ZGV8 mode for the 3-5 composite plate, with λ = 36, 49 and 35 μm, respectively.

## 4. THE ELECTROACOUSTIC COUPLING COEFFICIENT

The electroacoustic coupling coefficient $K^2$ is a measure of the IDTs electrical to acoustic energy conversion efficiency in piezoelectric materials. In the ZnO/a-SiC composite plates, four electroacoustic coupling configurations can be obtained by placing the interdigital **T**ransducers at the substrate/film interface (sTf) or at the film surface (sfT), further including a floating metal electrode opposite the IDTs (sTfm and smfT), as shown in figure 7.

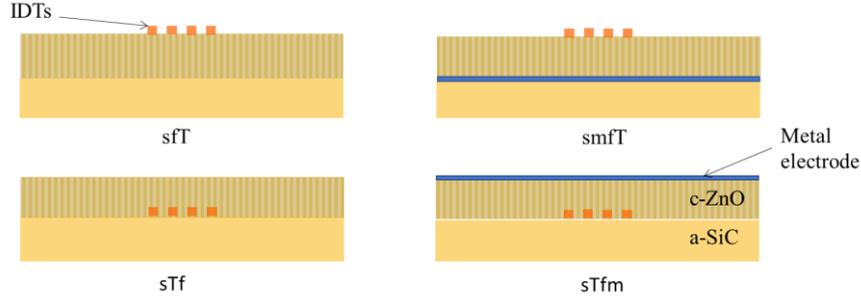

**Figure 7**: The four coupling configurations: sfT, smfT, sTf, and sTfm.

The $K^2$ of the ZGV modes was calculated by the following approximated formula, $K^2 = 2 \cdot [(v_f - v_m)/v_f]$, where $v_f$ and $v_m$ are the phase velocities of the mode for the free and electrically short-circuited surfaces of the ZnO film [28-30]. The $v_f$ and $v_m$ were calculated for different a-SiC and ZnO layers thicknesses using McGill software [31]; the ZnO and a-SiC were assumed to be lossless, and their material constants (mass density, elastic, piezoelectric, and dielectric constants) were extracted from references 23 and 27. The $K^2$ has dispersive characteristics and it is highly affected by the electrical boundary conditions. Figure 8 shows the $K^2$ vs the ZnO thickness curves of the ZGV2 mode, for the four coupling configurations, at fixed a-SiC layer thickness (5 μm thick). Each ZnO thickness corresponds a different wavelength: λ ranges from 20 to 50 μm for the 0.5-5 to the 5-5 ZGV2 plate. The sTfm and smfT configurations are the most efficient between the four: their $K^2$ reaches 3.9 and 3.3% for the 5-5 plate. The $K^2$ of the four configurations starts to decrease for further increase in the ZnO layer thickness.

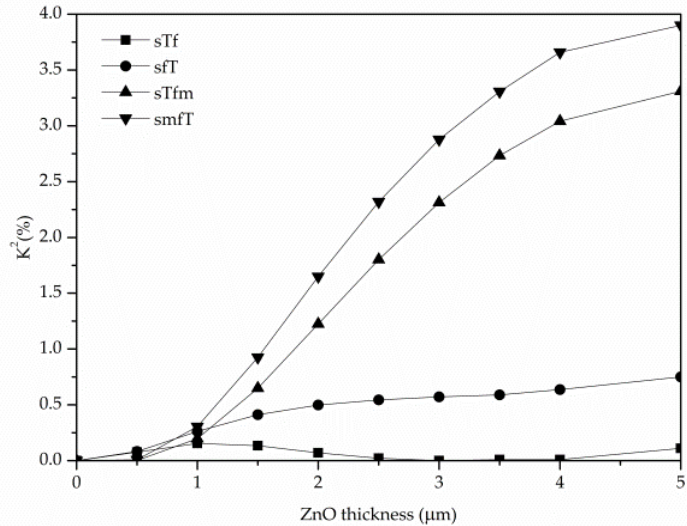

**Figure 8.** The $K^2$ dispersion curves of the ZGV2 mode of the four coupling configurations; the a-SiC layer thickness is fixed (5 μm thick).

As an example, figure 9 shows the admittance vs frequency curves of the ZGVR based on the ZGV2 of the 5-5 composite plate, for the four coupling configurations; the inset shows the effective electromechanical coefficients $k^2_{eff} = (\frac{f_p^2 - f_s^2}{f_p^2})$ of the four configurations. The smfT and sTfm configurations provide a $k^2_{eff}$ much higher than the sfT and sTf configurations; $f_s$ and $f_p$ are the series and parallel frequencies of the ZGVR extracted by the admittance frequency spectrum (the frequencies corresponding to the maximum conductance and maximum resistance [32]). The $k^2_{eff}$ is larger than $K^2$ since the present simulation accounts for the electrodes parameters (material constants and finite thickness) and the ZnO acoustic loss [33].

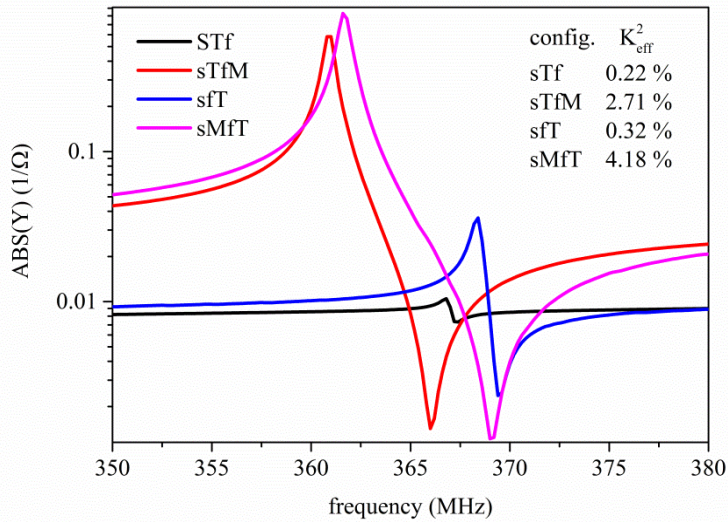

**Figure 9.** The absolute admittance vs frequency curves of the ZGVR of 5-5 composite plate, for the four coupling configurations.

Figure 10 shows the $K^2$ dispersion curves of the qS$_0$ mode, for the four coupling configurations: the a-SiC thickness is fixed (5µm) while the ZnO thickness, i.e. the graph's abscissa, varies in the same range exploited for the ZGV2 of figure 8. Both the ZGV2 and qS$_0$ modes can be excited on the same plate (with equal λ) with good and comparable K$^2$, mostly for the smfT and sTfm configurations. The qS$_0$ is a *propagating* mode, as opposed to the ZGV2, thus the design of a one port S$_0$-based LWR will include an IDT and two reflectors. Moreover, since the phase velocity of the qS$_0$ mode is significantly lower than that of the ZGV2, the resonant frequency of the qS$_0$-LWR will be much lower than that of the ZGVR.

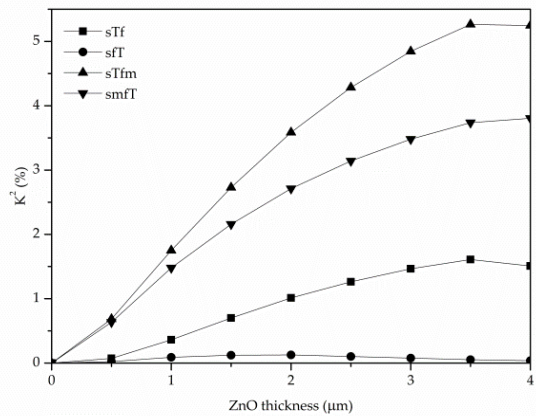

**Figure 10.** The K$^2$ vs the ZnO layer thickness curves of the S$_0$ mode, for the four coupling configurations and for fixed a-SiC thickness (5 μm thick).

Figure 11 shows, as an example, the *K*$^2$ dispersion curves of the ZGV5 for fixed a-SiC thickness (5 μm) and variable ZnO layer thickness, for the four coupling configurations.

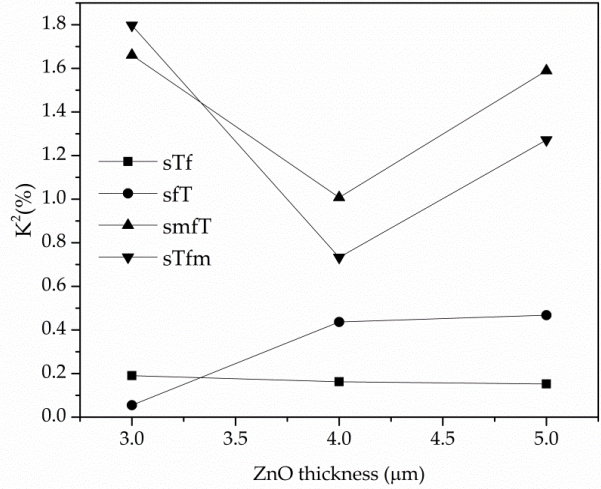

**Figure 11.** The *K*$^2$ dispersion curves of the ZGV5 mode of the four coupling configurations.

The K$^2$ of the ZGV5 is much lower than that of the ZGV2 but it shows a similar trend while increasing the piezoelectric layer thickness.

### 5. THE ZGV2 PRESSURE SENSOR

The acoustic energy of a ZGV mode is concentrated under the IDT but, if an external pressure is applied that induces the bending of the membrane, the resonator frequency is expected to change. The investigation of this feature can be exploited for sensing applications or for evaluating the resonator stability under variable environmental pressure. By following the calculation procedure outlined in reference 14, we investigated the major factors determining the pressure sensitivity of the ZGVR based on mode 2 of the 1-1 composite plate for the smfT configuration: the 1-1 plate was chosen as it corresponds the highest K$^2$, according to figure 8. The 1-1 ZGV2 and 5-5 ZGV2 are characterized by a different wavelength value (λ = 10 and 50 μm, respectively) but equal thickness-to-wavelength ratio (H$_{tot}$/λ = 2/10 = 0.2 and H$_{tot}$ = 10/50 = 0.2) which corresponds to the highest K$^2$ value, as shown in figure 8; the 1-1 ZGV2, besides having a K$^2$ equal to that of the 5-5 ZGV2, it has the additional technological advantage to require thinner layers and thus smaller sized device.

When a membrane is exposed to an external pressure, it resulted strained and the internal pressure changes. As a consequence, the material constants of the membrane change as well as the thickness and the geometrical dimensions of the membrane, thus resulting in a resonance frequency shift. The contributions to the pressure sensitivity of the ZGVRs due to the dependence on the pressure of the elastic constants, the lateral and vertical strains were studied. 2D and 3D FEM Comsol Multiphysics analysis has been performed to determine the pressure sensitivity of the ZGVR by two-steps simulations: 1. 3D stationary study of mechanical deflection of the membrane with symmetric boundary conditions, under uniform differential pressure; 2. 2D eigen-frequency study of a single pair of IDT at the H/λ corresponding to the ZGV2, with continuity boundary conditions. Figure 12 shows the schematic of the ZGV-based pressure sensor; the topological design parameters of the device are listed in table 1.

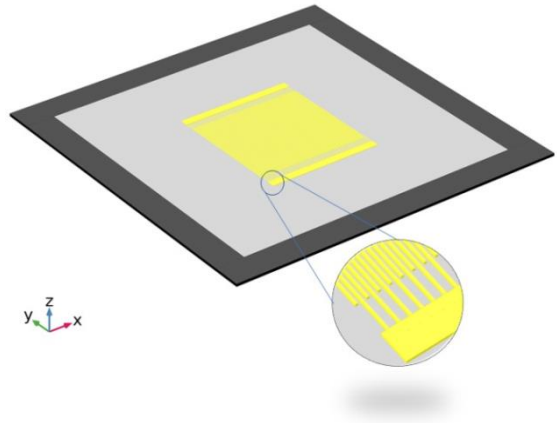

**Figure 12.** The schematic of the suspended membrane: the silicon frame is black, the membrane is gray; the IDT is yellow.

**Table 1.** The topological design parameters of the plate.

| Design parameter | value |
|---|---|
| Acoustic wavelength, $\lambda$ | 10 μm |
| ZnO film thickness, $h_{ZnO}$ | 1 μm |
| a-SiC film thickness, $h_{a-SiC}$ | 1 μm |
| Device aperture | 40 $\lambda$ |
| Number of finger pairs | 40 |
| Membrane Size | 1000 x 1000 μm$^2$ |
| IDT Area, AIDT | 400 x 400 μm$^2$ |
| Al electrodes thickness | 50 nm |

*5.1. 3D stationary study of mechanical deflection of the membrane*

Figure 13 shows the total displacement of the 1-1 composite plate under 1 kPa uniform pressure difference: a variable pressure is supposed to be applied to the lower surface of the membrane while the upper one is maintained at fixed (ambient) pressure. The vertical axis of the membrane deformation is scaled up by the factor of 10 for graphical reason, otherwise the deformation could not be clearly observed. Due to the symmetry of the problem, just a quarter of the membrane, with applied symmetric boundary conditions, is sufficient for the analysis. As it can be seen, most of the displacement is concentrated in the center of the membrane.

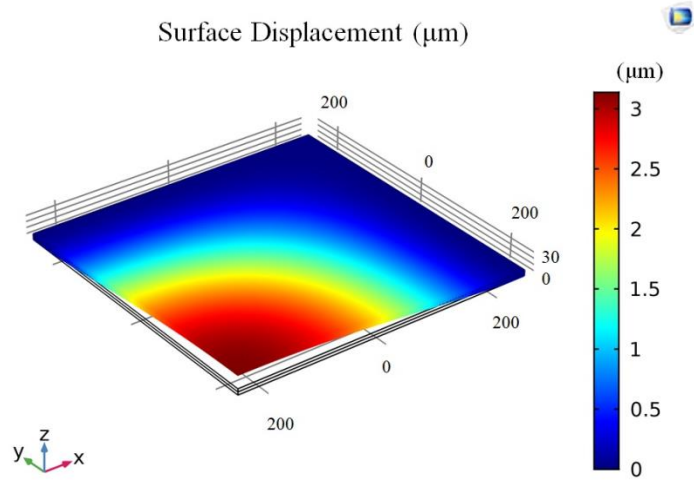

**Figure 13.** The total displacement of the 1-1 composite membrane under 1 kPa ambient pressure difference. The vertical axis of the membrane deformation is scaled up by the factor of 10.

The FEM model was used to determine the strain ($S_{xx}$, $S_{yy}$ and $S_{zz}$), stress ($T_{xx}$, $T_{yy}$ and $T_{zz}$) and internal pressure $P_{int} = -(T_{xx} + T_{yy} + T_{zz})/3$, at the interface of the two layers, as well as at the free surface of each layer. Figures 14a-c show $S_{xx}$, $S_{zz}$ and $P_{int}$ vs the distance from the center of the membrane. Each figure shows three curves: the solid and dash curves are referred to the free surface of the ZnO and a-SiC layer, and the dot-dash curves are referred to the interface between the two layers, respectively. The gray area represents half the IDT area, $A_{IDT}/2$: inside this area the $S_{xx}$, $S_{zz}$ and $P_{int}$ are almost constant.

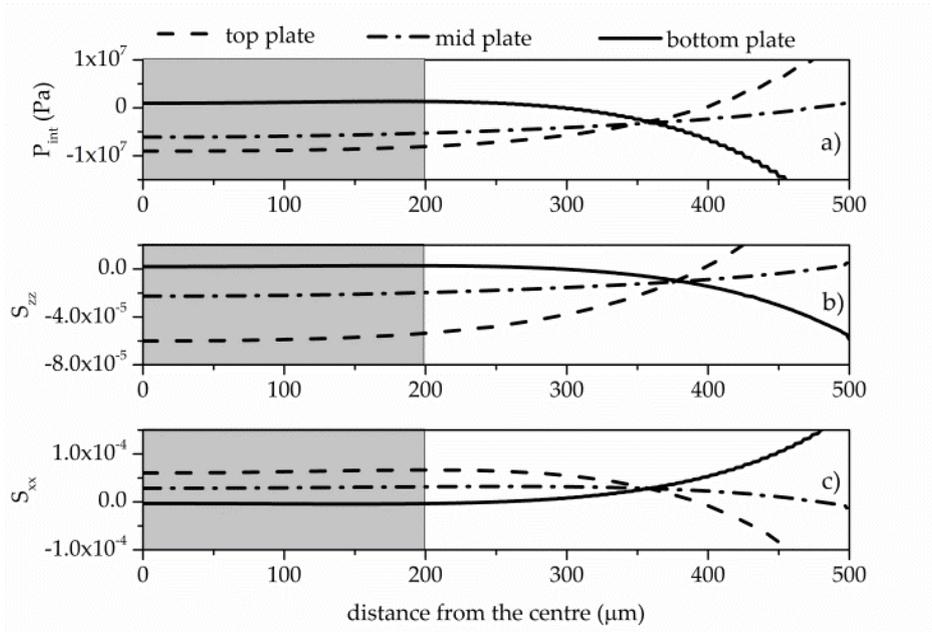

**Figure 14.** a) The internal pressure *Pint*, b) the $S_{zz}$ and c) $S_{xx}$ strain components vs the distance from the centre of the membrane; the solid and dash curves are referred to free surface of ZnO and a-SiC

layer, and the dot-dash curves are referred to the interface between the two layers, respectively. The gray area represents half the IDT area, A$_{IDT}$/2.

The volume average values of the strain and internal pressure, $\overline{S_{xx}}$, $\overline{S_{zz}}$ and $\overline{P_{int}}$, were derived inside the ZnO and the a-SiC layers, underneath the IDT area: the data obtained are listed in table 2.

**Table 2.** The internal pressure, and the $\overline{S_{xx}}$ and $\overline{S_{zz}}$ strain mean values of the 1-1 ZnO/a-SiC plate subjected to 1kPa uniform constant differential pressure.

| layer | $\overline{P_{int}}$ (MPa) | $\overline{S_{xx}}$ (ppm) | $\overline{S_{zz}}$ (ppm) |
|---|---|---|---|
| ZnO | -6.22 | 41.368 | -41.256 |
| a-SiC | -2.93 | 10.444 | -6.278 |

The $\overline{S_{xx}}$, $\overline{S_{zz}}$ and $\overline{P_{int}}$ represent the layers elongation, the thickness change and the internal pressure the plate is subjected to, due to the applied differential pressure. As it can be seen from table 2, the 1 kPa pressure induces a remarkable internal pressure (MPa order of magnitude).
The $\overline{S_{xx}}$, $\overline{S_{zz}}$ and $\overline{P_{int}}$ were then calculated for different external pressure values in the 1 to 10 kPa range.

*5.2. 2D eigen-frequency study*

The eigen-frequency study of the ZGV2 for the 1-1 composite plate was performed for three *perturbed* conditions:
  1. the IDT wavelength was assumed to be equal to

$$\lambda_{pert} = \lambda(1 + \overline{S_{xx}^{ZnO}}) \qquad (1)$$

under the hypothesis that the IDT is positioned onto the free surface of the piezoelectric layer, and that the wavelength is not affected by the $\overline{S_{xx}^{a-SiC}}$; the two layers thickness and material constants are assumed to be unaffected by the applied differential P.
  2. The thickness of each layer was assumed to be equal to

$$h_{ZnO}^{pert} = h_{ZnO}(1 + \overline{S_{zz}^{ZnO}}) \qquad (2)$$

and

$$h_{a-SiC}^{pert} = h_{a-SiC}(1 + \overline{S_{zz}^{aSiC}}). \qquad (3)$$

The wavelength and the materials constants are assumed to be unaffected by P.
  3. The two layers elastic constants and mass density were changed according to their pressure dependence

$$c_{ij}^{pert} = c_{ij} + (\partial c_{ij}/\partial P) \cdot P_{int} \qquad (4)$$

and

$$\rho_{pert} = \rho + (\partial \rho/\partial P) \cdot P_{int}; \qquad (5)$$

the thickness of each layer and the wavelength were assumed to be unaffected by P.

The a-SiC and ZnO pressure derivatives of the elastic constants [34, 35] $c_{ij}$ and mass density ρ are listed in table 3.

**Table 3**: The pressure derivatives of the mass density and elastic constants of ZnO and a-SiC.

| material | Pressure derivative of material constants | | |
|---|---|---|---|
| a-SiC | $\partial c_{11}/\partial P = 3.49$ | $\partial c_{12}/\partial P = 4.06$ | $\partial \rho/\partial P = 16.06$ (kg/m³GPa) |
| ZnO | $\partial c_{11}/\partial P = 3.8$ | $\partial c_{12}/\partial P = 5.2$ | $\partial c_{13}/\partial P = 4.7$ |
| | $\partial c_{33}/\partial P = 3.7$ | $\partial c_{44}/\partial P = -0.53$ | $\partial \rho/\partial P = 37.8$ (kg/m³GPa) |

The Murnaghan equation of state

$$P = \left(\frac{B}{B'}\right) \cdot \left[\left(\frac{V_0}{V_{pert}}\right)^{B'} - 1\right] \quad (6)$$

was used to calculate the w-ZnO relative mass density change $\rho_0/\rho$ caused by the applied pressure P, where $\rho$ and $\rho_0$ are the pressure-perturbed and unperturbed mass density, $B$ (142.6 GPa) and B' (3.6) are the ZnO bulk modulus and its pressure derivative, $B'= \partial B/\partial P$; $V_{pert}$ and $V_0$ are the perturbed and equilibrium volumes [36].
The Murnaghan equation of state for isotropic elastic solidi [37]

$$P = a \cdot (f + 5 \cdot f^2) \quad (7)$$

was used to evaluate the pressure derivative of the a-SiC mass density, with $f = 0.5 \cdot \left\{\left(V_0/V\right)^{2/3} - 1\right\}$ and $a = 3 \cdot c_{12} + 2 \cdot c_{44}$.

The pressure derivatives of the mass density for both the ZnO and a-SiC layers were also calculated by 3D FEM stationary study. The ZnO and a-SiC mass density, as a function of $P_{int}$, was deduced by the plate volume change $V_0/V_{pert} = \rho/\rho_0$, being $L_x$, $L_y$ and $L_z$ the equilibrium membrane sides along the axis system, $V_0 = L_x \cdot L_y \cdot L_z$ the unperturbed volume, $V_{pert} = V_0 \cdot (1 + S_{xx}) \cdot (1 + S_{yy}) \cdot (1 + S_{zz})$ the perturbed volume. The a-SiC and ZnO mass density evaluated by the two methods for different pressure values are very close, as shown in figure 15. The discrepancy between the data obtained with the two methods is from $7 \cdot 10^{-3}$ to about $3 \cdot 10^{-2}$ kg/m³ for ZnO, and from $7 \cdot 10^{-4}$ to $8.7 \cdot 10^{-3}$ kg/m³ for a-SiC, respectively, in the studied pressure range.

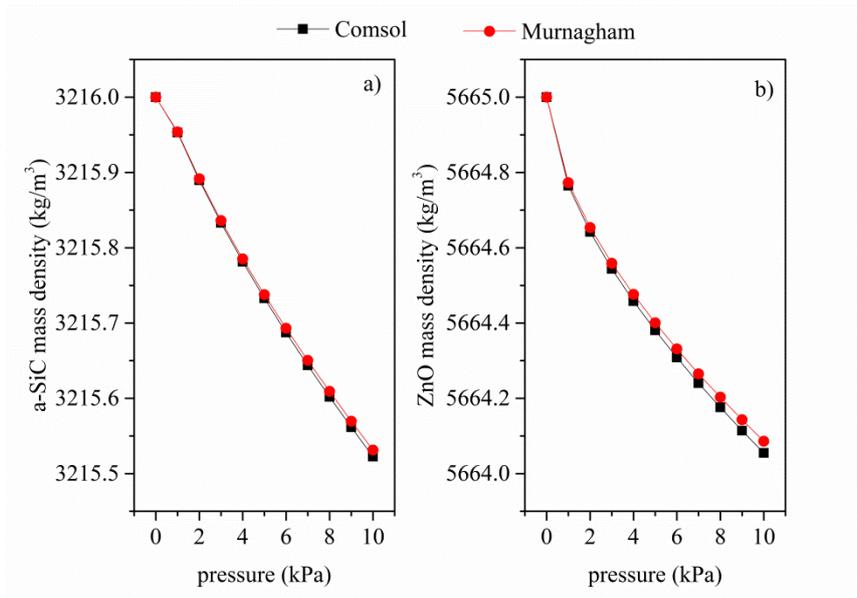

**Figure 15**: The mass density vs pressure curves of a) a-SiC and b) ZnO; the red circle and black square points were calculated by the Murnagham equation and by Comsol simulation, respectively.

Figure 16 shows the relative resonant frequency shifts of the ZGV2, $\Delta f/f_0$, induced by $\overline{S_{xx}}$, $\overline{S_{zz}}$ and $\overline{P_{int}}$ vs the applied differential pressure; $\Delta f = f_{pert} - f_0$, $f_0 = 1841.754$ MHz is the unperturbed resonant frequency, and $f_{pert}$ is the pressure-perturbed resonant frequency.

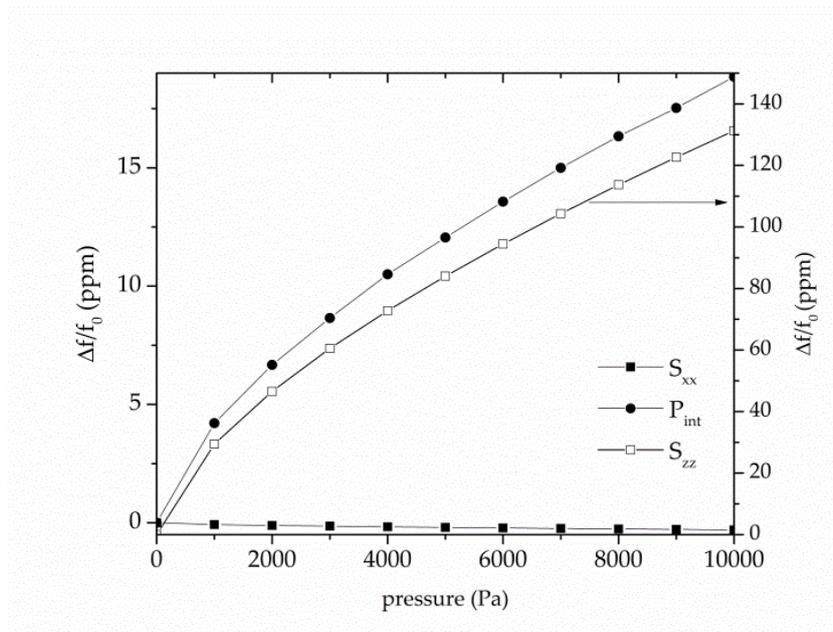

**Figure 16**: The $\overline{S_{xx}}$, $\overline{S_{zz}}$, $\overline{P_{int}}$ -induced ZGVR relative frequency changes vs the applied differential pressure for the 1-1 composite plate.

From figure 16 it appears evident that the ZGV2 relative frequency shifts induced by the changes in the membrane length, thickness and internal pressure are quite different. The effect provided by the $\overline{S_{zz}}$ changes is positive since the membrane thickness decreases under the applied pressure, and the wave velocity (and hence the resonant frequency) consequently increases. This effect is dominant over the others since the ZGV highly dispersive behavior ensures a large sensitivity to any thickness changes. The contribution related to $\overline{S_{xx}}$ is negative as the wavelength increases under the applied pressure, and the resonant frequency consequently decreases; it is also marginal since the mode corresponds to the zero group velocity dispersion. The effect provided by the $\overline{Pint}$ changes is positive and smaller than that provided by $\overline{S_{zz}}$, as opposed to the results shown in ref. 14 where the $\overline{Pint}$ contribution is dominant and the pressure-induced mass density changes are not accounted for. In the present simulation the a-SiC and ZnO pressure-induced mass density changes are accounted for: the mass density contribution has the effect to lower the elastic constants contribution to the mode velocity increase.

Figure 17 shows the *sum* of the relative frequency shifts shown in figure 16, $\left(\Delta f/f_0\right)_{SUM} = \left(\Delta f/f_0\right)_{Sxx} + \left(\Delta f/f_0\right)_{Szz} + \left(\Delta f/f_0\right)_{Pint}$, vs the differential pressure. Figure 17 also shows the *total* frequency shift $\left(\Delta f/f_0\right)_{total}$ vs the differential pressure: the ordinate was calculated by changing, at the same time, the membrane size (thickness and length) and the two layers material constants (mass density and elastic constants).

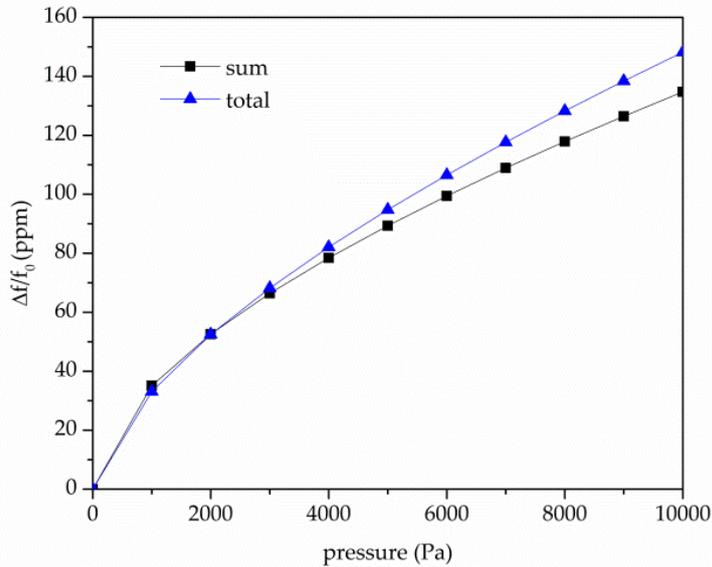

**Figure 17**: The $\left(\Delta f/f_0\right)_{sum}$ and the $\left(\Delta f/f_0\right)_{total}$ vs the pressure curves of the ZGV2-based sensor on the 1-1 composite plate.

In the 4 to 10 kPa pressure range, the two curves of figure 17 can be linearly fitted, showing a sensor sensitivity, i.e. the curve slope, of 9.34 and 10.98 ppm/kPa, respectively. As the two curves are very similar, then we can argue that the net response of the ZGVR-pressure sensor can be approximated, in the studied pressure range, as the sum of the responses caused by the different contributions.
As an example, figure 18 shows the ZGV2 absolute admittance vs frequency curves for the 1-1 plate at three differential pressure values (0 Pa, 5 kPa and 10kPa).

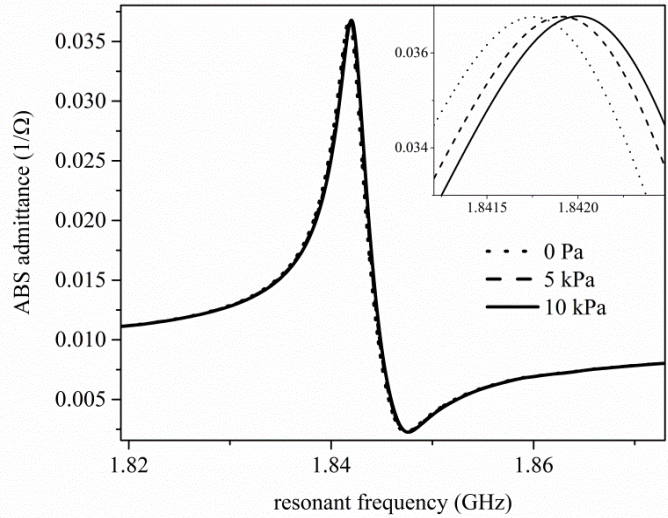

**Figure 18**: The ZGV2 absolute admittance vs frequency curves of the 1-1 plate, for the sfT configuration; the pressure is the running parameter.

The non-linear behavior of the ZGVR-based pressure sensor, shown in figure 17, is in agreement with that of the high-frequency ($f_0$ = 10.77 GHz) one port SAWR-based pressure sensor implemented on AlN(300 nm)/diamond(20 μm) suspended substrate, when a pressure variation in the range from 0 to 100 kPa is applied, as described in reference 20. The rough side of the diamond free standing layer was glued with an epoxy adhesive to an alumina bulk substrate (few hundreds of μm thick) with a hole in the middle. The pressure underneath the hole was kept constant at atmospheric pressure, while a variable pressure was applied to the upper surface of the membrane. The SAWR-based sensor has a pressure sensitivity (0.31 ppm/kPa) lower than that of the ZGVR-based sensor, but it has a smaller surface area (the periodicity of the finger pairs is 800 nm, the number of periods was 100 for both the IDT and the reflectors).
In reference 21 a fully implantable wireless pressure sensor is presented that works in the 0 to 26.66 kPa pressure range and shows a sensitivity of about 12 ppm/kPa. The sensor device consists in a 5.3 by 4 mm thin quartz diaphragm on which a one-port SAWR has been implemented that works at about 868 MHz. Since the total space occupied by an implantable sensor also includes the antenna, it is important to minimize its size. The small size of the sensor based on ZGV2 1-1, as well as its compatibility with integrated circuit technology and its sensitivity to low pressures, make it a suitable candidate for the design of implantable devices.
In reference 22 the effect of the diaphragm shape (circular and rectangular) on the pressure sensitivity of an AlN/SOI-based SAWR sensor was investigated in the 0 to 1724 kPa pressure range, and an improved sensitivity from 0.0055 to 0.025 ppm/kPa was found. This result suggests us to study the effects of the membrane shape on the pressure sensitivity of the ZGV-based sensor, as well as the optimization of the IDT design parameters, also including the presence of reflectors to compensate a possible deviation of the theoretical thicknesses from the calculated ones.

## 6. Conclusions
The propagation characteristics of the ZGV Lamb modes along c-ZnO/a-SiC composite plates have been modelled for different ZnO and a-SiC layer thicknesses. The phase and group velocity, and the $K^2$ of four coupling configurations have been theoretically studied specifically addressing the design of enhanced-coupling, microwave frequency electroacoustic devices that are reliable to fabricate by conventional sputtering technologies and microlithography technique. Quite good $K^2$ corresponds to both the $qS_0$ and the ZGV2, for the same-thickness plate, but the former has a resonant frequency significantly lower than that of the ZGVR, and requires a larger device area including the IDT and two reflectors. The pressure sensitivity of the ZGV2 resonator was studied for the 1-1 plate subjected to a uniform differential pressure varying in the 1 to 10 kPa range. A ZGVR sensitivity of about 9 ppm/kPa was predicted in the 4 to 10 kPa range where the relative frequency change vs the pressure curve can be linearly fitted.

The present study was performed to demonstrate proof of concept of ZGVRs in pressure sensing applications. The c-ZnO/a-SiC-based ZGV2 sensors are proven to achieve remarkable performances (high sensitivity and enhanced coupling efficiency) that are important prerequisite for the design of future devices to be used in the context of chemical, biological and physical quantities detection. Further studies are in progress to improve the device performances, based on the IDT design parameters (such as single electrode and multiple-split electrodes) and the membrane shape.

**Acknowledgment**
This project has received funding from the European Union's Horizon 2020 research and innovation programme under the Marie Sklodowska-Curie Grant Agreement No. 642688.